\documentclass[onecolumn]{jpsj3}
\usepackage{color}
\usepackage{txfonts}
\usepackage{epstopdf}
	\usepackage{fancyhdr}
\renewcommand{\thispagestyle}[1]{} 

\title{Critical temperature of site-diluted spin-1/2 systems with long-range ferromagnetic interactions}

\author{Karol Sza{\l}owski\thanks{E-mail: kszalowski@uni.lodz.pl}, and Tadeusz Balcerzak}
\inst{Department of Solid State Physics, Faculty of Physics and Applied Informatics, \\University of {\L}\'{o}d\'{z},\\ ulica Pomorska 149/153, 90-236 {\L}\'{o}d\'{z}, Poland} 

\abst{In the paper the Pair Approximation (PA) method for studies of the site-diluted spin-1/2 systems of arbitrary dimensionality with the long-range ferromagnetic interactions is adopted. The method allows to take into account arbitrary anisotropy of the interactions in the spin space, so it is not limited to purely Ising couplings. Within this approach, the  Gibbs free energy is obtained, which allows to derive all the further interesting thermodynamic properties. In particular, we obtain an equation for the critical temperature of the second-order phase transitions for the model in question. In the study we focus our attention on the systems with ferromagnetic interactions decaying with the distance according to the power law $J(r)\propto r^{-n}$. We discuss the dependence of the critical temperature on the concentration of magnetic component and the index $n$ for selected one-, two- and three-dimensional lattices. We confirm the absence of the critical concentration for a diluted magnet with infinite 
interaction range. In the regime of the low concentrations of magnetic component, we find a non-linear increase of the critical temperature with the concentration in the form of $T_{c}\propto p^{n/d}$, depending on the system dimensionality $d$ and the index $n$, whereas $n>d$. }

\kword{Ising model, Heisenberg model, critical temperature, long-range interactions} 

\begin{document}
\maketitle

\section{Introduction}

The studies of the systems with the long-range interactions constitute a challenging contemporary problem in statistical physics \cite{Bouchet}. The important part of the studies concerns the systems with the so called 'strong long-range interactions' \cite{Bouchet}, which term denotes the couplings decaying with the distance slow enough to cause a failure of extensivity which is the basis for formulation of thermodynamics. However, there is a wide class of systems in which such a behaviour does not emerge and usually formulated thermodynamics is an appropriate and valuable tool for their characterization. Within this field, a range of magnetic systems attracted considerable attention focusing mainly on low dimensions \cite{Joyce1,Fisher,Mermin,Nolting,Barati,Barati1,Barati2,Monroe,Monroe2,Monroe3,Diep,Diep2,Curilef,Nakano,Nakanob,Tomita,Pires,Pires2,Khoo,Hamedoun,Laflorencie,Luijten,Sandvik1,Pacobahyba,Yusuf,Zhu,Pino,Cavallo1,Cosenza1,Vassiliev2001a,Vassiliev2001b,Vassiliev2001c,Nakano1,Nakano2,Cavallo2,Cavallo4,
Campana}. This selection is generally restricted to magnets in which interactions are of constant sign, thus not leading to magnetic frustration with a plethora of intriguing consequences. Let us mention that the studies of magnetic systems with the site dilution and long-range couplings seem to be rather rare and this subject is principally mentioned only in the context of spin glasses and scaling relations \cite{Chowdhury,Cyrot}.

Let us present a brief motivation for studies of diluted magnetic systems with long-range interactions provided by some recent experimental works. One can instance the progress in growth and characterization of a highly promising dilute magnetic semiconductor (Ga,Mn)N, which encourages the interest in three-dimensional ferromagnets with the long-range interactions, for this substance attracts rising interest in the context of potential room-temperature ferromagnetism \cite{Sylwia1,Sylwia2,Sylwia3,Dietl1,Dietl2}. In this compound, a non-linear dependence of the critical temperature on magnetic Mn dopant concentration has been found experimentally for low Mn content, and such behaviour has been attributed to a ferromagnetic long-range superexchange mechanism \cite{Dietl1,Dietl2,Sylwia2}. What is more, the unique properties of indirect Ruderman-Kittel-Kasuya-Yosida interaction in graphene (see e.g. \cite{RKKY,Duffy}) also promote theoretical understanding of two-dimensional magnets with the long-range coupling (e.g. \cite{Fabritius1,Qi}). 

Despite the development and use of simulational Monte Carlo methods for the systems with the long-range interactions \cite{Fabritius1,Dietl2,Fukui,Watanabe,Sasaki}, there is still room for analytic studies. 
However, the problem turns out to be complex and, up to now, no complete thermodynamic method, which goes beyond the Molecular Field Approximation (MFA), has been proposed. In order to fill the gap,
the present work describes the thermodynamics of the site-diluted systems with spins $1/2$ interacting ferromagnetically by means of  the long-range coupling, using analytical method based on the Pair Approximation (PA). The PA method is superior to MFA from the point of view of the systematic hierarchy of Cluster Variational Methods (CVM) \cite{Kikuchi,Morita, Katsura}. These methods have been originally developed for the nearest-neighbour (NN) interactions. However, the application of CVM for larger clusters: for instance, in triangle or square approximation, in the presence of the long-range interaction does not seem to be possible in practice. Nevertheless, it turns out that in the frame of CVM reduced to the PA the problem of long-range interactions is still tractable. The usefulness of the PA method follows from the fact that, in contrast to MFA, it takes into account the spin-pair correlations and can be applied to low-dimensional and disordered magnets \cite{Balcerzak2}. Moreover, this method 
yields the Gibbs free-energy from which all thermodynamic quantities can be calculated.

In this paper, within the PA method, the equation for the critical (Curie) temperature for the system in question has been obtained. Attention is being focused on a specific form of long-range interactions, namely decaying with the distance according to the power law. For such a coupling the dependence of the critical temperature on the concentration of magnetic atoms for various anisotropy parameters characterizing the coupling has been illustrated and discussed.

\section{Theoretical model}

The Hamiltonian of a spin-1/2 site-diluted ferromagnet with the long-range interactions can be written in the following form:
\begin{equation}
\mathcal{H}=-\sum_{i,\,j}^{}{J\left(r_{ij}\right)\left[\Delta\left(S^{x}_{i}S^{x}_{j}+S^{y}_{i}S^{y}_{j}\right)+S^{z}_{i}S^{z}_{j}\right]\xi_{i}\xi_{j}}-h\sum_{i}^{}{\xi_{i}S^{z}_{i}},
\label{eq1}
\end{equation}
where $J_{k}=J\left(r_{ij}\right)>0$ is the ferromagnetic exchange integral between two spins $i$ and $j$, the distance between which amounts to $r_{ij}=r_{k}$. It is assumed that one of the spins is the $k$-th nearest-neighbour of the other one, i.e. this spin belongs to the $k$-th coordination zone around the central one and the set of radii $r_{k}$ for $k=1,2,\dots$ characterizes fully a given crystalline lattice. The parameter $0\leq \Delta \leq 1$ is the anisotropy of  interaction in the spin space, and is assumed to be independent on the distance between interacting spins. $\Delta=0$ corresponds to Ising interaction, while $\Delta=1$ is the isotropic Heisenberg coupling. The site dilution is introduced by means of the occupation number operators $\xi_{i}$, for which the configurational average $\left\langle \xi_i\right\rangle=p$ yields the concentration of the magnetic atoms. The external magnetic field is denoted by $h$. Since the interaction is long-ranged, the summation in the Hamiltonian extends for all site pairs of the considered crystalline lattice. 

In order to describe the thermodynamics of the model in question, the Pair Approximation method is extended to be capable of treating the systems with the long range interactions. 
The method is based on the cumulant expansion technique for the free energy \cite{Katsura}, which constitutes a systematic approach used in the frame of CVM. In the PA only the first- and second-order cumulants are taken into account, and the higher-order cumulants are neglected. This corresponds to the assumption that only the single-site and pair cluster energies contribute to the total energy. The spin-spin interactions within each cluster pair are taken exactly. The molecular fields in which the clusters are embedded play a role of variational parameters. These parameters can be self-consistently determined from the condition that the Gibbs energy in equilibrium must achieve a minimum. Moreover, the magnetizations calculated basing on single sites and on pairs must be equal, which is imposed by a consistency condition.

The PA method has been previously applied to the extensive studies of various magnetic systems with the interactions limited to nearest neighbours \cite{Balcerzak2,Balcerzak3,Szalowski1,Szalowski2,Szalowski3} and has been exhaustively described there; therefore, only a brief scheme is presented here. 

The quantum state of a spin is described by means of the following density matrices:
\begin{equation}
\hat{\rho}^{i}=e^{\beta G^{(1)}}\exp\left[\beta\left(\Lambda+h\right)S^{i}_{z} \right]
\label{rho1}
\end{equation}
for a single spin at site $i$ and
\begin{equation}
\hat{\rho}^{ij}=e^{\beta G^{(2)}_{k}}\exp \left\{\, \beta\,  J_{k}\left[\Delta\left(S^{x}_{i}S^{x}_{j}+S^{y}_{i}S^{y}_{j}\right)+S^{z}_{i}S^{z}_{j}\right]
+\left(\Lambda'_{k}+h\right)\left(S^{i}_{z}+S^{j}_{z}\right) \right\} 
\label{rho2}
\end{equation}
for a pair of spins at sites $i$ and $j$. Here, $J_{k}$ is the interaction for the $k$-th coordination zone of the given crystalline lattice and $\beta=1/\left(k_{\rm B}T\right)$.

In the present formulation the total Gibbs energy per site, averaged over magnetic component configurations, $\left\langle G\right\rangle _{r}=\left\langle\left\langle \mathcal{H}\right\rangle\right\rangle_{r}-\left\langle S \right\rangle_{r}T$, can be expressed in the following form:
\begin{equation}
\frac{\left\langle G\right\rangle_{r}}{N}=\frac{1}{2}p\sum_{k=1}^{\infty}{\left[z_{k}pG^{(2)}_{k}-2\left(z_{k}p-1\right)G^{(1)}\right]},
\label{Gibbs}
\end{equation}
where $z_{k}$ is the number of lattice sites belonging to the $k$-th coordination zone. The single-site and pair Gibbs energy terms are:
\begin{equation}
G^{(1)}=-k_{\rm B}T\ln \left[2\cosh\left(\beta \frac{\Lambda+h}{2}\right)\right]
\label{Gibbs1}
\end{equation}
and
\begin{eqnarray*}
G^{(2)}_{k}&=&-k_{\rm B}T\ln \left\{2\exp\left(\beta\frac{J_{k}}{4}\right)\cosh\left[\beta\left(\Lambda'_{k}+h\right)\right]\right.\\&&\left.+2\exp\left(-\beta\frac{J_{k}}{4}\right)\cosh\left(\beta\frac{J_{k}\Delta}{2}\right)\right\},
\label{Gibbs2}
\end{eqnarray*}
respectively. 

The parameter $\Lambda$ has the interpretation of a molecular field acting on a single spin and originating from all the spins in its environment. The analogous parameter $\Lambda'$ denotes a molecular field acting on a selected pair of spins, one of them being a $k$-th nearest neighbour of the other. Both parameters can be further expressed using the variational parameters $\lambda_j$ which constitute molecular fields acting on given spin and resulting from its interaction with an $j$-th nearest neighbour spin. Therefore we can write:
\begin{equation}
\Lambda=p\sum_{l=1}^{\infty}{z_{l}\lambda_{l}}
\label{Lambda}
\end{equation}
and
\begin{equation}
\Lambda'_{k}=\Lambda-\lambda_{k}=\sum_{l=1}^{\infty}{\left(pz_{l}-\delta_{kl}\right)\lambda_{l}}.
\label{Lambdaprim}
\end{equation}

The variational minimization of the Gibbs energy with respect to $\lambda_j$ is performed with a set of constraints in the form of $\mathrm{Tr}_{i}\left(\hat{\rho}^{i} S^{i} \right)=\frac{1}{2}\mathrm{Tr}_{ij}\left[\hat{\rho}^{ij}\left( S^{i}+S^{j}\right) \right]$, which impose a condition that the magnetization for a given lattice site is the same when calculated using either a single-site or a pair density matrix. Such a procedure leads to the self-consistent set of equations in the form of:
\begin{eqnarray}
&&\tanh\left[\frac{1}{2}\beta \left(\Lambda + h\right)\right]=\nonumber\\
&&=\frac{e^{\frac{1}{4}\beta J_{k}}\sinh\left[\beta\left(\Lambda'_{k}+h\right)\right]}{e^{\frac{1}{4}\beta J_{k}}\cosh\left[\beta\left(\Lambda'_{k}+h\right)\right]+e^{-\frac{1}{4}\beta J_{k}}\cosh\left(\frac{1}{2}\beta J_{k}\Delta\right)},
\label{equations}
\end{eqnarray}
where $k=1,2,\dots$ numbers the subsequent coordination zones. After plugging in  Eq. (\ref {equations}) the formulas (\ref{Lambda}) and (\ref{Lambdaprim}) the set of equations for $\lambda_{j}$ variables is finally obtained.

The solution to the infinite set of self-consistent equations (\ref{equations}) allows the Gibbs energy to be determined  and hence the thermodynamic behaviour of the system can be completely characterized. Further thermodynamic quantities of interest can be obtained as appropriate derivatives of the Gibbs energy with respect to its natural variables.

It should be emphasized here that the Gibbs energy, which has been constructed from the enthalpy $\left\langle\left\langle \mathcal{H}\right\rangle\right\rangle_{r}$ (i.e., the mean value of the Hamiltonian containing interaction with the external field $h$) and the entropic part $\left\langle S \right\rangle_{r}T$, in general, is a function of three parameters: $h$, $T$ and $N$. Since we are using the canonical ensemble with $N=const.$, only the temperature $T$ and the external field $h$ are (intensive) thermodynamic parameters for which the Gibbs energy can be treated as a thermodynamic potential. As these two variables can easily be controlled in the experiment, they appear to be very convenient in magnetism.

The present paper focuses on the critical temperature of the second-order phase transition for ferromagnetic system. 

It is worth mentioning here that in the system with spin $S=1/2$ and solely ferromagnetic NN interactions we do not expect to obtain the 1st order phase transitions. Such discontinuous phase transitions may occur  when the competitive interactions (often introduced for higher spins), or magnetic frustration take place, which is not our case.
For the continuous phase transitions the derivation is presented in details in Appendix A. 

Within this approach, we obtain the following equation for the critical (Curie) temperature $T_{c}$:
\begin{equation}
p\sum_{k=1}^{\infty}{z_{k}\left[1-\exp\left(-\frac{1}{2}\beta_{c}J_{k}\right)\cosh\left(\frac{1}{2}\beta_{c}J_{k}\Delta\right)\right]}=2,
\label{TC}
\end{equation}
where $\beta_{C}=1/\left(k_{\rm B}T_{C}\right)$. This equation will serve as a basis for numerical calculations, the results of which are discussed in the following section. Let us mention here that an usual Molecular Field Approximation leads to the following formula for the Curie temperature:
\begin{equation}
k_{\rm B}T^{MFA}_{C}=\frac{1}{4}p\sum_{k=1}^{\infty}{z_{k}J_{k}},
\label{TCMFA}
\end{equation}
which is insensitive to the interaction anisotropy in the spin space.

Let us remark  that for a specific case of Ising couplings limited only to nearest-neighbour spins, i.e., when $J_1>0,J_2=J_3=\dots=0$ and $\Delta=0$, we can solve Eq.~(\ref{TC}) to obtain the expression for the critical temperature in the form $k_{\rm B}T_{c}/J_1=1/\left\{2 \ln\left[pz/\left(pz-2\right)\right]\right\}$, which agrees with the results previously reported in Refs.~\cite{Balcerzak2003,Balcerzak3,Szalowski1,Balcerzak2}. On the other hand, for Heisenberg couplings with $\Delta=1$ we obtain $k_{\rm B}T_{c}/J_1=1/\ln\left[pz/\left(pz-4\right)\right]$ \cite{Balcerzak2}.

We should also mention  that the validity of our approach, the outcome of which is Eq.~(\ref{TC}), is limited to such interactions $J\left(r\right)$, for which the sum in Eq.~(\ref{TC}) is convergent and all thermodynamic quantities resulting from the used formulas (like the Gibbs energy per site) are finite. This implies that the interaction should decay fast enough with the distance between magnetic moments. 

In order to illustrate the critical temperatures resulting from the equation (\ref{TC}), let us assume for further calculations a specific form of distance dependence of couplings between magnetic moments. For this purpose we will select a power-law decay of the  interaction, in the form of:
\begin{equation}
J_{k}=J_{1}\left(r_{k}/r_{1}\right)^{-n},
\label{Jr}
\end{equation}
($k=1,2,\dots$),
where $J_1$ and $r_1$ are the coupling energy and distance between nearest neighbours for a given lattice. The exponent $n>0$ characterizes the power decay. This form of distance dependence of the coupling is known as 'magnetic Gr\"{u}neisen law' and has been postulated in \cite{Bloch}. Moreover, such a dependence is also used for interpretation of the  experimental data \cite{Coffman}. Let us mention that such a formula is an empirical one and is applied to both ferro- and antiferromagnetic interactions. All the results presented below will be normalized to the parameter $J_1$ (i.e., NN interactions) setting the energy scale.

One of the interesting issues is the character of dependence of the Curie temperature on the concentration of magnetic component $p$. In particular, the range of small $p$ is of special interest. In this regime the detailed structure of the underlying crystalline lattice is not expected to be important, leaving only the dependence on the dimensionality $d$ of the lattice. As a consequence, a continuous approximation can be applied to the Eq. (\ref{TC}), the details of which are presented in Appendix B. The resulting formula for the critical temperature for small $p$ is:
\begin{equation}
k_{\rm B}T_{c}=J_{1}\left[\left(1+\Delta\right)^{d/n}+\left(1-\Delta\right)^{d/n}\right]^{n/d}\left(\frac{r_{1}}{\Omega_0^{1/d}}\right)^{n} \frac{1}{2} \left(\frac{\omega_{d} \Gamma\left(1-\frac{d}{n}\right)}{4d}\right)^{n/d}\,p^{n/d}.
\label{TCeq}
\end{equation}
for $d=1,2,3$; the coefficient $\omega_{d}=2$ for $d=1$, $\omega_{d}=2\pi$ for $d=2$ and $\omega_{d}=4\pi$ for $d=3$.
The presented result is valid only for the exponent $n>d$, for the distance dependence of the interaction given by (\ref{Jr}). $\Omega_0$ denotes the volume/area/length (depending on the dimensionality) of the system per site. $\Gamma\left(x\right)$ is Euler gamma function. The condition $n>d$ is used to guarantee the convergence of the total energy and associated quantities, including convergence of the sum in Eq.~(\ref{TC}).

The most important finding from Eq.~(\ref{TCeq}) is that the critical temperature is no longer proportional to $p$, as in MFA (Eq. (\ref{TCMFA})). Instead, it varies in a non-linear way, proportionally to $p^{n/d}$. Let us mention that such a dependence, $T_{C}\propto  p^{n/d}$, can be inferred from the scaling analysis presented briefly in Refs.~\cite{Chowdhury,Cyrot}. This kind of non-linear dependence remains unmodified by various values of interaction anisotropy $\Delta$ in the  spin space. 

Let us observe, along the lines of the discussion in Ref. \cite{Fabritius1}, that for a diluted system with concentration of magnetic component equal to $p$, the average distance between the impurities $r_{av}$ amounts to $r_{av}=\left(\Omega_0/p\right)^{1/d}\propto p^{-1/d}$. The interaction energy between impurities at this distance is $J_{av}\propto p^{n/d}$. Therefore, for very low concentration $p$, the critical temperature is governed by the coupling between magnetic impurities at average distance. On the other hand, for high concentration $p\to 1$, the critical temperature tends to vary linearly with $p$.

Another interesting problem concerns the existence of the finite critical concentration $p_{c}$ below which the critical temperature vanishes. Such a critical concentration has been found within the PA method for the  diluted systems with interaction limited to the nearest neighbours only \cite{Balcerzak2}. In the presented case, from  Eq. (\ref{TC}) (or from its alternative form (\ref{eqB1})), in the limit $T_c \to 0$ and $J\left( r\right)>0$ for all $r<\infty$, we can obtain $\displaystyle p_{c}= \frac{2}{\sum_{k=1}^{\infty}{z_{k}}}$ for the Ising model ($\Delta=0$) and $\displaystyle p_{c}= \frac{4}{\sum_{k=1}^{\infty}{z_{k}}}$ for the pure Heisenberg system ($\Delta=1$). For NN interactions only we have $\sum_{k=1}^{\infty}{z_{k}}=z_1$ and the critical concentrations reduce to these reported in Ref. \cite{Balcerzak2}. From the above formulas it is clear that $p_{c}\to 0$ if the interaction does not vanish totally for any finite distance between impurities, as then $\sum_{k=1}^{\infty}{z_{k}}\to \infty$. Therefore, the critical temperature for the interacion given by Eq.(\ref{Jr}) is always nonzero for any finite $p$.
It is worthy mentioning that the physical meaning of the critical concentration is not only connected with vanishing of the Curie temperature but also, from the structural point of view, indicates the percolation threshold in dilute systems.

\section{Numerical results and discussion}

\begin{figure}
\includegraphics[scale=0.26]{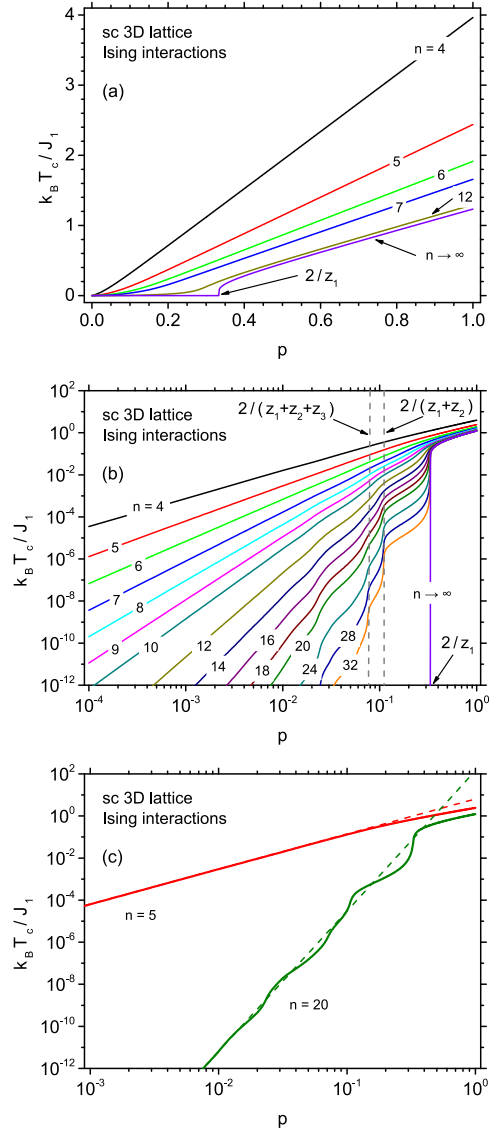}
\caption{(Color online) Dependence of critical temperature on magnetic component concentration for various indexes $n$ on linear scale (a) and double logarithmic scale (b). Comparison of the numerical solution of Eq.~(\ref{TC}) (solid lines) with low-concentration analytical approximation Eq.~(\ref{TCeq}) (dashed lines) for two different indexes $n$ in double logarithmic scale (c). Exchange integral between nearest neighbours $J_1$ is fixed. 3D simple cubic (sc) lattice is considered, with $z_1 = 6$, $z_2 = 12$, $z_3 = 8$. Ising couplings (with $\Delta = 0$) are assumed.}
\label{fig:fig1}
\end{figure}

In Fig.~\ref{fig:fig1}(a) we present the dependence of the normalized critical temperature on the concentration of magnetic component, plotted for three-dimensional simple cubic (sc) lattice with $z_1=6$ nearest neighbours, in a linear scale. The presence of long-range Ising interactions is assumed. The dependencies for various values of index $n$ are shown, starting from $n=4$. It is evident that for the values of $p$ larger than $2/z_{1}$ the dependencies for all the values of $n$ are linear in their character, and their slope  is decreasing with increasing $n$. For the lowest value of $n$ the curve remains almost linear in the whole range of concentrations. However, for larger $n$ values, a kink emerges close to $p=2/z_1$. For $n\to \infty$ we reproduce the results for Ising model with nearest-neighbour interactions only, i.e. the critical concentration  $p_{c}=2/z_1=1/3$ is present, below which $T_{c}=0$.

\begin{figure}
\includegraphics[scale=0.26]{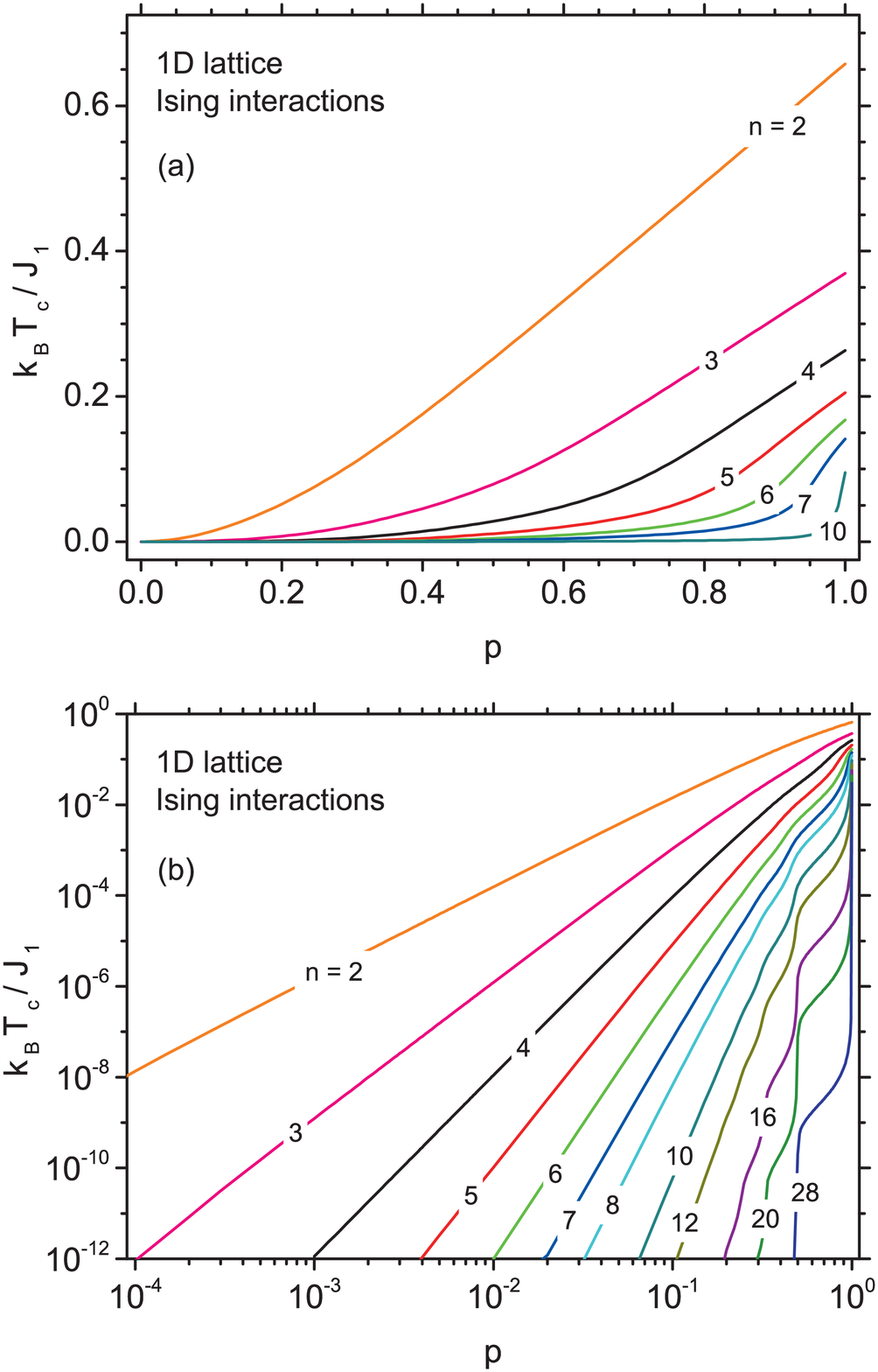}
\caption{(Color online) Dependence of critical temperature on magnetic component concentration for various indexes $n$  on linear scale (a) and double logarithmic scale (b). Exchange integral between nearest neighbours $J_1$ is fixed. 1D lattice (linear chain) is considered, with $z_1 = 2$. Ising couplings (with $\Delta = 0$) are assumed.}
\label{fig:fig2}
\end{figure}

\begin{figure}
\includegraphics[scale=0.26]{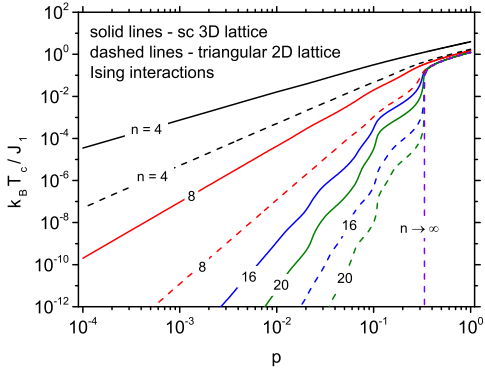}
\caption{(Color online) Dependence of critical temperature on magnetic component concentration for various indexes $n$  on double logarithmic scale. Exchange integral between nearest neighbours $J_1$ is fixed. The two lattices with $z_1=6$ are compared: 3D sc lattice (solid lines) and 2D triangular lattice (dashed lines). Ising couplings (with $\Delta = 0$) are assumed.}
\label{fig:fig3}
\end{figure}

\begin{figure}
\includegraphics[scale=0.26]{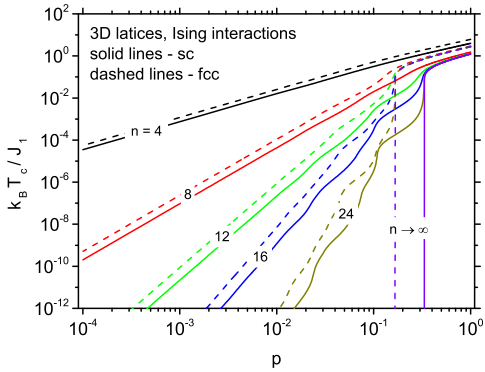}
\caption{(Color online) Dependence of critical temperature on magnetic component concentration for various indexes $n$  on double logarithmic scale. Exchange integral between nearest neighbours $J_1$ is fixed. The two 3D lattices are compared: sc lattice with $z_1 = 6$ (solid lines) and fcc lattice with $z_1 = 12$ (dashed lines). Ising couplings (with $\Delta = 0$) are assumed.}
\label{fig:fig4}
\end{figure}

\begin{figure}
\includegraphics[scale=0.26]{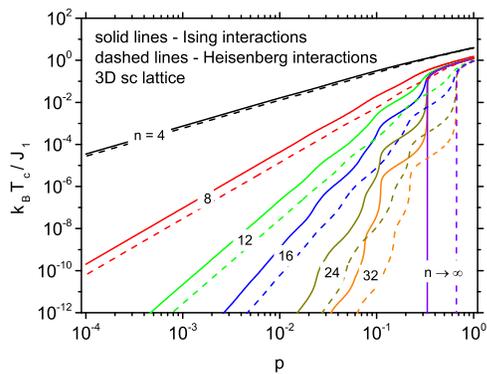}
\caption{(Color online) Dependence of critical temperature on magnetic component concentration for various indexes $n$  on double logarithmic scale. Exchange integral between nearest neighbours $J_1$ is fixed. The two models: Ising  (solid lines) and isotropic Heisenberg (dashed lines) are compared for 3D sc lattice.}
\label{fig:fig5}
\end{figure}

\begin{figure}
\includegraphics[scale=0.26]{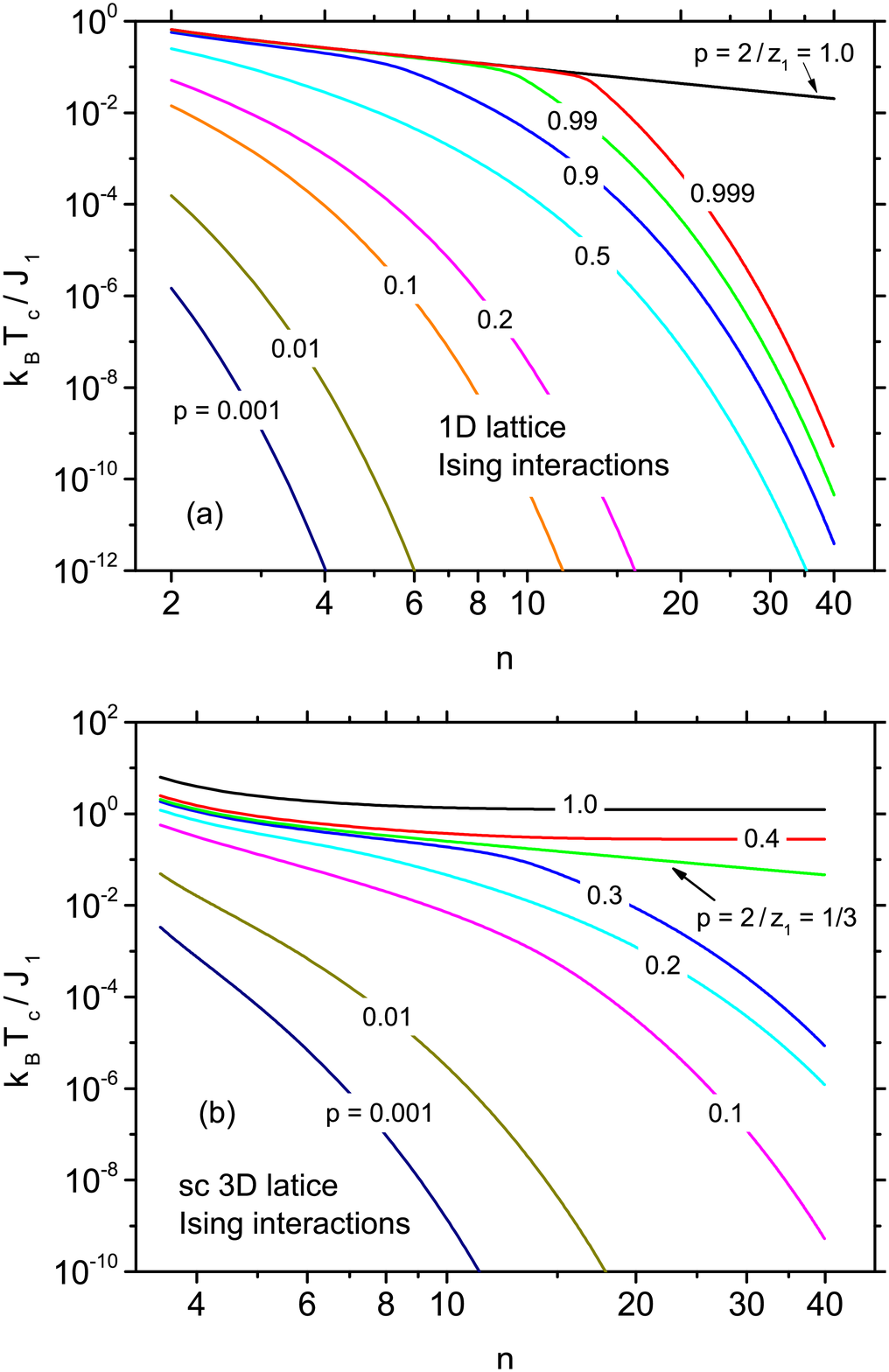}
\caption{(Color online) Dependence of critical temperature on index $n$ for various concentrations of magnetic component $p$, on double logarithmic scale. Exchange integral between nearest neighbours $J_1$ is fixed. Ising couplings (with $\Delta = 0$) are assumed, for (a) 1D lattice (linear chain) with $z_1 = 2$; (b) 3D sc lattice with $z_1 = 6$.} 
\label{fig:fig6}
\end{figure}

It is instructive to present the same dependence on the double logarithmic scale, as plotted in Fig.~\ref{fig:fig1}(b). From such presentation it is visible that for the lowest concentrations $p$, the dependencies of $T_{c}$ vs. $p$ become linear, which is a sign of power-law dependence. This observation is in accordance with Eq.~(\ref{TCeq}), where $T_{c}\propto p^{n/d}$ is predicted. Moreover, it is evident that the slope of the curves on the double logarithmic scale is increasing with the increase of $n$ in the  low concentration range. 
In order to better illustrate the comparison between the analytical and numerical results, in Fig.~\ref{fig:fig1}(c) we plot two selected solutions of the general Eq.~(\ref{TC}) (solid curves) together with their analytical approximations presented by Eq.~(\ref{TCeq}) (dashed lines) for two different values of exponent $n$. One can see that for sufficiently low concentration $p$ the analytical approximation given by Eq.~(\ref{TCeq}) is fully consistent with the numerical solution of the full equation for critical temperature (\ref{TC}).

For large values of $n$, corresponding to a considerably fast decrease of the coupling with the distance, a series of subsequent kinks is visible in Fig.~\ref{fig:fig1}(b). The first one corresponds to $p=2/z_1$, while the positions of the other correspond to $p=2/\left(z_1+z_2\right)$, $p=2/\left(z_1+z_2+z_3\right)$, etc. The positions of the above mentioned kinks are indicated in the plot with dashed vertical lines. 
According to the discussion of the critical concentration, and the formulas presented at the end of previous Section, these values correspond to the critical concentrations, which would appear when the interactions were cut off at the first, second, third, etc., coordination zone, respectively. Since such cutting-off does not take place when $n< \infty$, the critical temperature does not fall to zero at those values; instead, only the rapid decrease of critical temperature takes place and a noticeable kink is formed at the curve.
When $n\to \infty$, the behaviour of $T_{c}$ is convergent to the behaviour of the  Ising model with interaction only between nearest neighbours (and the value of $p_c=1/3$ in this case simultaneously corresponds to the first kink for all other curves).

One can notice that a similar plot has been presented in the work Ref.~\cite{Fabritius1}, where the two-dimensional graphene has been considered with antiferromagnetic couplings decaying according to the law $J\left(r\right)\propto r^{-3}$. Quantum Monte Carlo results for isotropic Heisenberg model show the proportionality of the critical temperature to $p^{3/2}$, which is also in accord with our results (Eq.~\ref{TCeq}). Moreover, for larger concentrations $p$ a linear dependence of the critical temperature on magnetic impurity concentration is found, as in the presented results. Let us also observe that a lack of kink on the dependence of the critical temperature vs. $p$ in the results of Ref.~\cite{Fabritius1} is also in qualitative agreement with what we obtain for low values of index $n$ (for the case considered in Ref.~\cite{Fabritius1}, $n=3$ and $d=2$).  

Let us also present analogous dependencies calculated for one-dimensional lattice (chain), which are shown in Fig.~\ref{fig:fig2}(a) and (b), on the linear and double logarithmic scale, respectively. Contrary to 3D sc lattice, for 1D system there is nonlinear regime of $T_{c}$ for large values of $p$. One can see that the first kink in Fig.~\ref{fig:fig2}(b) emerges at $p=1$, and when index $n$ increases the critical temperature at this kink quickly drops to zero. In the limiting case, when $n \to \infty$, only the NN interaction $J_1$ remains ($J_k$ =0 for $k\ge2$ on the basis of Eq~(\ref{Jr})). Then, we found that the critical temperature tends to zero for all concentrations, including $p=1$. This result is in accordance with the exact solution for the linear Ising chain with NN interactions, where no phase transition occurs at non-zero temperatures. In the range of small concentrations of magnetic atoms and finite $n$, the features are 
rather similar to the ones present in 
the previous case (i.e. the presence of further kinks and power-law dependence of $T_{c}$ on $p$ when $p\to 0$ can be seen in Fig.~\ref{fig:fig2}(b)). 

Fig.~\ref{fig:fig3} presents a comparison of the results obtained for two crystalline lattices with the same number of nearest neighbours ($z_1=6$), but of different dimensionality, namely for  3D sc lattice and 2D triangular (tr) lattice. The critical temperatures were calculated for Ising couplings and plotted as a function of concentration $p$ on a double logarithmic scale, for selected values of $n$. 
The calculated $T_{c}$ is always lower for a 2D system than for a 3D system. The difference in the critical temperatures tends to vanish for the increasing index $n$ in the range of large concentrations $p$. This reflects the fact that for large $n$ the most important role is played by the interaction with nearest neighbours, the number of which is equal for both selected lattices. In the limit of $n\to\infty$ the results fall onto the same curve which is predicted from the application of the Pair Approximation to a diluted magnet with nearest-neighbour interactions only. 
In this case the first kink appears at $p_1^{sc}=p_1^{tr}=2/z_1=1/3$. 

The further kinks for these curves are connected with the next coordination zones. Next two coordination numbers for sc lattice are $z_2=12$ and $z_3=8$. This leads to the concentrations corresponding to the second and third kinks: $p_2^{sc}=2/\left(z_1+z_2\right)=1/9$ and  $p_3^{sc}=2/\left(z_1+z_2+z_3\right)=1/13$. For the triangular lattice the first three coordination numbers are equal: $z_1=z_2=z_3=6$. Thus, the second and the third kinks appear at the values: $p_2^{tr}=1/6$ and $p_3^{tr}=1/9$, respectively. It is worthy noticing that the second kink for sc lattice and the third kink for tr lattice appear at the same concentration $p_2^{sc}=p_3^{tr}=1/9$. This coincidence can be visible, for example,  on the curves with index $n=20$. It means that the critical concentration for 3D sc lattice with the first- and second-neighbour interactions is the same as the critical concentration for 2D tr lattice, where the interactions up to the third coordination zone are taken into account. 
A remarkable feature of the critical temperature dependencies on $p$ is the difference in slope for a low concentration range between the  curves plotted for both systems of unequal dimensionality $d=2$ and $d=3$. This behaviour is in concert with Eq.~\ref{TCeq}.

Fig.~\ref{fig:fig4} illustrates the results for two lattices of the same dimensionality - sc 3D lattice and fcc 3D lattice. In this case it is visible that the critical temperature is higher for fcc lattice where the density of sites is greater. However, the slope of the dependence of $T_{c}$ vs. $p$ on the double logarithmic scale is the same in low $p$ range for both lattices, since their dimensionality is the same. The positions of kinks observable on both curves are different since the numbers $z_{k}$ are mostly unequal for these crystalline lattices. In particular, we have $z_1=6$ for sc while $z_1=12$ for fcc lattice. This difference causes a different limiting critical temperature behaviour for both lattices when $n\to\infty$.

The effects of the interaction anisotropy are studied in Fig.~\ref{fig:fig5}, where the results of critical temperature calculation are compared for sc 3D lattice with either Ising or isotropic Heisenberg couplings. It is evident that the critical temperatures are lowered by switching from the anisotropic to isotropic coupling. This effect is least remarkable for low $n$ and becomes gradually more and more pronounced when $n$ increases. The slope on the double logarithmic scale is the same for low $p$ (see Eq.~\ref{TCeq}) and does not depend on the interaction anisotropy $\Delta$. However, the limiting high-$n$ behaviour differs, for the Pair Approximation predicts $T_{c}=0$ below $p_{c}=2/z_{1}$ for $\Delta<1$ and below $p_{c}=4/z_{1}$ for $\Delta=1$ (in agreement with Ref.~\cite{Balcerzak2}).

It can also be of interest to study the dependence of the critical temperature on index $n$ for some fixed values of concentration $p$. Such plots are presented in Fig.~\ref{fig:fig6}, for Ising couplings on 1D lattice (a) and on 3D sc lattice (b). A double logarithmic scale is used. For 1D lattice, the critical temperature drops with increasing $n$ and the tendency is stronger for higher $n$ values. In this case the drop in $T_{c}$ is not limited by a non-zero value. When the concentration $p$ increases, the range of slower drop of $T_{c}$ emerges for lower $n$ and for $p$ close to 1 this range is significant. For $p=1$ the dependence is different, because only a slow linear-like drop of the critical temperature for increasing $n$ is visible. Let us observe that $p=1$ is a limiting case for 1D lattice for which $z_1=2$ and thus $p_{c}=2/z_1=1$. It means that if $n \to \infty$ then $T_c \to 0$, in agreement with the exact result for the Ising chain with NN interactions. Somewhat similar behaviour can be seen in Fig.~\ref{fig:fig6}(b) for 3D sc lattice. When $p<p_{c}=2/z_1=1/3$, the behaviour of $T_{c}$ (unlimited, fast drop) is 
analogous to one observed in Fig.~\ref{fig:fig6}(a). However, in the range of concentrations $p_{c}=1/3<p\leq 1$ qualitatively different dependence of $T_{c}$ vs. $n$ is seen. Namely, after some initial decrease, the critical temperature tends to the limiting value predicted by the Pair Approximation for a diluted magnet with the nearest-neighbours coupling only. The separating line for $p=p_{c}=1/3$ corresponds to  the slow linear-like decrease in the critical temperature.

\section{Final remarks and conclusion}

In the paper the Pair Approximation method for  spin-1/2 systems with the  long-range couplings of ferromagnetic character and random site dilution has been applied. In particular, we found the equation for the critical (Curie) temperature with  the interaction anisotropy $\Delta$ taken into account. For the interesting case of interactions varying with the distance between spins like $J\left(r\right)\propto r^{-n}$ a limiting formula for critical temperature (valid in the limit of low concentration $p$) has been derived. This formula shows that the critical temperature varies non-linearly with the concentration of magnetic atoms, namely $T_{c}\propto p^{n/d}$, where $d$ is the dimensionality of the considered system. This finding differs qualitatively from the Mean Field Approximation prediction, where $T_{c}\propto p$ for any interaction and dimensionality. The prediction of our method is in agreement with scaling arguments \cite{Chowdhury,Cyrot} where the same proportionality has been found. The result is also in accord with some Quantum Monte Carlo calculations for honeycomb lattice \cite{Fabritius1} ($d=2$) with spin $S=1/2$ and interaction of the type $J\left(r\right)\propto r^{-3}$ ($n=3$). Namely, the result found in Ref.~\cite{Fabritius1} is $T_{c}\propto p^{3/2}$ for $p\leq 0.2$. There is also a strong experimental evidence that $T_c$ for very diluted magnets with long-range interaction is non-linear. For instance, the experiments performed on 3D dilute magnetic semiconductors  (DMS) $\rm {Ga}_{1-p}\rm {Mn}_{p}\rm{N}$ \cite{Sylwia2,Sylwia3, Dietl1, Dietl2} gave the result $T_{c}\propto p^{2.2}$ for $p \leq 0.1$. The scaling law with a similar exponent ($p^{1.9}$) is obeyed by spin-glass freezing temperature in Co-based II-VI DMS \cite{Twardowski, Swagten, Shand}. Also in a wide class of Mn-based DMS power-law dependence of freezing temperature on magnetic ion concentration is confirmed \cite{Galazka}. In our plots the dependence 
of the critical temperature on the concentration of magnetic atoms for various lattices of different dimensionality has been illustrated.

In our work we focused our attention on the phase transition temperature calculation. The critical behaviour in the vicinity of the phase transition has not been studied; however, it has been known that the critical exponents in the PA method are the same as in the Landau theory, i.e., given by MFA. For the regular lattices such classical critical exponents present an approximation. It has also been shown that the PA method gives exact results when is applied for the Bethe lattices with NN interactions \cite{Tucker}.

The differences between the Ising and Heisenberg models in the PA method can be noticed through different phase transition temperatures  and different critical concentrations. In particular, for NN interactions only (when $n \to \infty$) the critical concentration obtained here for the Ising model is $p_c=2/z_1$, whereas for the Heisenberg model $p_c=4/z_1$. This means that 1D Ising chain with $z_1=2$ is nonmagnetic for non-zero temperatures, and 2D Heisenberg system with $z_1=4$ is also nonmagnetic (in accordance with Mermin-Wagner theorem \cite{Mermin}). Unfortunately, for NN interaction the PA method is not able to distinguish between 2D triangular lattice with $z_1=6$ and 3D simple cubic lattice. However, such lattices are distinguishable for the long-range interaction (Fig.~\ref{fig:fig3}).

As far as the NN interactions are concerned within the PA method, a difference between the Ising and Heisenberg models can also be found in the low-temperature behaviour of magnetic susceptibility. For instance, it has been found in Ref.\cite{Balcerzak3} that the susceptibility in the isotropic Heisenberg bilayer in the vicinity of $T=0$ diverges like $\propto 1/T$. One can suppose that such kind of behaviour  may also occur for the long-range  interactions; however, it needs more extended studies of all thermodynamic properties, which is beyond the scope of the present paper.

As far as the low-dimensional magnetism is concerned, we found that a non-zero critical temperature is found in all the systems where the interactions extend to infinity, provided $n>d$. This result is in accordance with theoretical predictions of several papers, for example: Quantum Monte Carlo method for 2D Heisenberg model \cite{Vassiliev2001c}, spherical model in 1D Ising system \cite{Joyce1},  one- and two-dimensional quantum Heisenberg model studied by spin wave theory \cite{Nakano1}, Green Function technique \cite{Nakano2} and Spectral Density method \cite{Cavallo2}.

Another interesting limit of interaction considered in literature is $n=0$, i.e., when the interactions extend to infinity and all of them have the same strength. Then, assuming $J_k=J_1/N$ (for the energy convergence), we obtain the Kac model \cite{Stanley}. That model has been solved exactly for the crystalline case giving the phase transition temperature and the molecular-field-like behaviour. However, in the case of dilution, we do not expect to obtain the non-zero critical concentration for the Kac model, similarly to MFA.

As for the context of the validity of our approach, let us once more put emphasis on the fact that our description is valid when the interaction decays appropriately fast with the distance (i.e., $n>d$ for $J\left(r\right)\propto r^{-n})$). Therefore, such a kind of 'long-range interactions' does not involve the systems for which the standard formulation of thermodynamics is not working properly: \cite{Bouchet} for example, due to failure of extensivity of some thermodynamic variables caused by a slow decay of interactions. As a consequence, the interactions we consider fall into the category of  the 'weak long-range interactions' according to classification in Ref.~\cite{Bouchet}. However, we are convinced that such a class of interactions is interesting; for example, from the modern magnetic systems point of view.

\begin{acknowledgments}
The computational support on Hugo cluster at Department of Theoretical Physics and Astrophysics, P. J. \v{S}af\'{a}rik University in Ko\v{s}ice is gratefully acknowledged.

This work has been supported by Polish Ministry of Science and Higher Education on a special purpose grant to fund the research and development activities and tasks associated with them, serving the development of young
scientists and doctoral students.
\end{acknowledgments}

\appendix
\section{Determination of the critical temperature}

The set of equations for the variational parameters takes the form of:
\begin{eqnarray}
&&\tanh\left[\frac{1}{2}\beta \left(\Lambda + h\right)\right]=\nonumber\\
&&=\frac{e^{\frac{1}{4}\beta J_{k}}\sinh\left[\beta\left(\Lambda'_{k}+h\right)\right]}{e^{\frac{1}{4}\beta J_{k}}\cosh\left[\beta\left(\Lambda'_{k}+h\right)\right]+e^{-\frac{1}{4}\beta J_{k}}\cosh\left(\frac{1}{2}\beta J_{k}\Delta\right)},
\label{eqA1}
\end{eqnarray}
where the values of the index $k=1,2,\dots$ number the subsequent coordination zones for the considered crystalline lattice. First, let us assume that the set of equations is truncated after $k_{max}$-th coordination zone, i.e. $k=1,\dots,k_{max}$. 

The variational parameters $\Lambda$ and $\Lambda'_{k}$ can be written as follows:
\begin{equation}
\Lambda=p\sum_{l=1}^{k_{max}}{z_{l}\lambda_{l}}
\label{eqA2}
\end{equation}
and
\begin{equation}
\Lambda'_{k}=\Lambda-\lambda_{k}=\sum_{l=1}^{k_{max}}{\left(pz_{l}-\delta_{kl}\right)\lambda_{l}}
\label{eqA3}
\end{equation}

The equations (\ref{eqA1}) can be linearized in the vicinity of the continuous phase transition, which yields:
\begin{equation}
\frac{1}{2}\beta\left[\Lambda-\frac{2}{1+e^{-\frac{1}{2}\beta J_{k}}\cosh\left(\frac{1}{2}\beta J_{k}\Delta\right)}\Lambda'_{k}\right]=0.
\label{eqA4}
\end{equation}

After substituting \ref{eqA2} and \ref{eqA3} into \ref{eqA4} we obtain the system of equations in the form:
\begin{equation}
\sum_{l=1}^{k_{max}}{M_{kl}\lambda_{l}=0},
\label{eqA5}
\end{equation}
with the matrix elements 
\begin{equation}
M_{kl}=pz_{l}-2\frac{pz_{l}-\delta_{kl}}{1+e^{-\frac{1}{2}\beta J_{k}}\cosh\left(\frac{1}{2}\beta J_{k}\Delta\right)}.
\label{eqA6-1}
\end{equation}

The equation for the critical (Curie) temperature of the continuous phase transition can be derived from the condition:

\begin{equation}
\det \left(M\right)_{kl}=0
\label{eqA6}
\end{equation}

By denoting:
\begin{eqnarray}
A_{l}&=&pz_{l}\qquad B_{k}=e^{-\frac{1}{2}\beta J_{k}}\cosh\left(\frac{1}{2}\beta J_{k}\Delta\right)-1\qquad\\ C_{k}&=&\frac{1}{1+e^{-\frac{1}{2}\beta J_{k}}\cosh\left(\frac{1}{2}\beta J_{k}\Delta\right)}\qquad\qquad\qquad,
\end{eqnarray}
we can write the matrix elements as follows:
\begin{equation}
M_{kl}=C_{k}\left(A_{l}B_{k}+2\delta_{kl}\right).
\end{equation}
Then, after some algebra, we obtain the expression for the determinant in the following form:
\begin{equation}
\det \left(M\right)_{kl} = 2^{k_{max}-1}\left(\prod_{k=1}^{k_{max}}{C_k}\right) \left(\sum_{k=1}^{k_{max}}{A_k B_k}+2\right),
\end{equation}
and the equation \ref{eqA6} for critical temperature is equivalent to
\begin{equation}
\sum_{k=1}^{k_{max}}{A_k B_k}+2=0,
\end{equation}
yielding finally:
\begin{equation}
p\sum_{k=1}^{k_{max}}{z_{k}\left[1-\exp\left(-\frac{1}{2}\beta_{c}J_{k}\right)\cosh\left(\frac{1}{2}\beta_{c}J_{k}\Delta\right)\right]}=2.
\end{equation}

Now, by assuming the limit $k_{max}\to\infty$ the  final result takes the form of:
\begin{equation}
p\sum_{k=1}^{\infty}{z_{k}\left[1-\exp\left(-\frac{1}{2}\beta_{c}J_{k}\right)\cosh\left(\frac{1}{2}\beta_{c}J_{k}\Delta\right)\right]}=2.
\label {eqA13}
\end{equation}

\section{Critical temperature dependence on magnetic component concentration for small concentrations}

The equation for the critical temperature \ref{eqA13} can be re-written as:
\begin{eqnarray}
&&\sum_{k=1}^{\infty}{z_{k}\left\{1-\exp\left[-\frac{\beta_{c}}{2}J_{k}\left(1+\Delta\right)\right]\right\}}\nonumber\\&&+\sum_{k=1}^{\infty}{z_{k}\left\{1-\exp\left[-\frac{\beta_{c}}{2}J_{k}\left(1-\Delta\right)\right]\right\}}=\frac{4}{p}
\label{eqB1}
\end{eqnarray}

Let us introduce the notation: $C_{\pm}\equiv J_{1}\left(1\pm\Delta\right)r_{1}^{n}/2$. Then, for the interactions $J_{k}\propto r_{k}^{-n}$, we get:

\begin{equation}
\sum_{k=1}^{\infty}{z_{k}\left[1-\exp\left(-\beta_{c}C_{+}r_{k}^{-n}\right)\right]}+\sum_{k=1}^{\infty}{z_{k}\left[1-\exp\left(-\beta_{c}C_{-}r_{k}^{-n}\right)\right]}=\frac{4}{p}
\label{B2}
\end{equation}
For $p\to 0$ we can replace summation over the coordination zones with integration over the volume/surface/length in the following way:
\begin{equation}
\sum_{k=1}^{\infty}{z_{k}\left[1-\exp\left(-\beta_{c} C_{\pm} r_{k}^{-n}\right)\right]} \to \frac{\omega_{d}}{\Omega_0}\int_{0}^{\infty}{r^{d-1}\left[1-\exp\left(-\beta_{c}C_{\pm} r^{-n}\right)\right]\,dr},
\end{equation}
where $\omega_{d}=2$ for $d=1$, $\omega_{d}=2\pi$ for $d=2$ and $\omega_{d}=4\pi$ for $d=3$.

It can be shown that for $n>d$ we get the result \cite{Mathematica}:
\begin{equation}
\int_{0}^{+\infty}{r^{d-1}\left[1-\exp\left(-\beta_{c}C_{\pm}r^{-n}\right)\right]\,dr}=\frac{1}{d}\left(\beta_{c}C_{\pm}\right)^{d/n} \Gamma\left(1-\frac{d}{n}\right),
\end{equation}
where $\Gamma\left(x\right)$ is the Euler gamma function. The condition $n>d$ is necessary to guarantee the convergence of the integrals and thus the finite value of the total energy of the system in question.

Using the above results we obtain from \ref{B2}:
\begin{equation}
\beta_{c}\left[\left(C_{+}^{d/n}+C_{-}^{d/n}\right)^{n/d}\right]=\left(\frac{4\Omega_0 d}{\omega_{d} \Gamma\left(1-\frac{d}{n}\right)}\right)^{n/d}\,p^{-n/d}
\end{equation}
for $d=1,2,3$.

Finally, the critical temperature can be expressed as follows:
\begin{equation}
k_{\rm B}T_{c}=J_{1}\left[\left(1+\Delta\right)^{d/n}+\left(1-\Delta\right)^{d/n}\right]^{n/d}\left(\frac{r_{1}}{\Omega_0^{1/d}}\right)^{n} \frac{1}{2} \left(\frac{\omega_{d} \Gamma\left(1-\frac{d}{n}\right)}{4d}\right)^{n/d}\,p^{n/d}.
\end{equation}
for $d=1,2,3$.

\end{document}